\documentclass{article}

\usepackage{arxiv}

\usepackage[utf8]{inputenc} % allow utf-8 input
\usepackage[T1]{fontenc}    % use 8-bit T1 fonts
\usepackage{hyperref}       % hyperlinks
\usepackage{url}            % simple URL typesetting
\usepackage{booktabs}       % professional-quality tables
\usepackage{amsfonts}       % blackboard math symbols
\usepackage{nicefrac}       % compact symbols for 1/2, etc.
\usepackage{microtype}      % microtypography
\usepackage{graphicx}
\usepackage{natbib}
\usepackage{doi}
\usepackage{times}
\usepackage{soul}
\usepackage{url}
\usepackage[small]{caption}
\usepackage{graphicx}
\usepackage{amsmath}
\usepackage{booktabs}
\usepackage{algorithm}
\usepackage{algpseudocode}
\usepackage{xcolor}
\usepackage[switch]{lineno}
\usepackage{longtable}
\usepackage{array}
\usepackage{algorithmicx}
\usepackage{multirow}
\usepackage{amssymb}
\usepackage{multirow, colortbl, xcolor}
\usepackage{booktabs}
\usepackage{listings}
\usepackage{array} % 用于定义新列类型
\usepackage{subcaption} % 引入子图宏包
\usepackage{enumitem} % 用于自定义列表格式
\newcolumntype{C}[1]{>{\centering\arraybackslash}p{#1}}
\newcolumntype{M}[1]{>{\centering\arraybackslash}m{#1}}
\usepackage{svg}
\usepackage{colortbl} % 确保加载colortbl包
\usepackage{xcolor}
\usepackage{siunitx}
\definecolor{headergray}{gray}{0.9} % 表头灰色
\definecolor{groupgray}{gray}{0.95} % 分组浅灰色
\definecolor{rowaltgray}{gray}{0.98} % 条纹背景（行）
\usepackage{fancybox}  % For fancy boxes
\usepackage{tabularx}
\usepackage{makecell}
\usepackage{xcolor}
\usepackage{algorithm}
\usepackage{algpseudocode}

\definecolor{lightgray}{gray}{0.95}
\definecolor{darkgray}{gray}{0.4}
\definecolor{purple}{rgb}{0.58,0,0.82}
\definecolor{blue}{rgb}{0.13,0.13,1}
\usepackage{geometry}

\title{AgentNet: Decentralized Evolutionary Coordination for LLM-based Multi-Agent Systems}

\date{} 					% Or removing it

\author{
\normalsize \textbf{Yingxuan Yang}\textsuperscript{1}\thanks{These authors contributed equally to this work.}, 
\normalsize ~\textbf{Huacan Chai}\textsuperscript{1}\footnotemark[1], 
\normalsize ~\textbf{Shuai Shao}\textsuperscript{1}, \\
\normalsize \textbf{Yuanyi Song}\textsuperscript{1}, 
\normalsize \textbf{Siyuan Qi}\textsuperscript{1}, 
\normalsize \textbf{Renting Rui}\textsuperscript{1}, 
\normalsize \textbf{Weinan Zhang}\textsuperscript{1,2}\thanks{Weinan Zhang is the corresponding author.}\\
\normalsize \textsuperscript{1}Shanghai Jiao Tong University, \normalsize \textsuperscript{2}SII\\
\normalsize \texttt{\{zoeyyx, fatcat, wnzhang\}@sjtu.edu.cn}
}

\hypersetup{
pdftitle={A template for the arxiv style},
pdfsubject={q-bio.NC, q-bio.QM},
pdfauthor={David S.~Hippocampus, Elias D.~Striatum},
pdfkeywords={First keyword, Second keyword, More},
}

\begin{document}
\maketitle

\begin{abstract} 
The rapid advancement of Large Language Models (LLMs) has catalyzed the development of multi-agent systems, where multiple LLM-based agents collaborate to solve complex tasks.   However, existing systems predominantly rely on centralized coordination, which introduces scalability bottlenecks, limits adaptability, and creates single points of failure.   Additionally, concerns over privacy and proprietary knowledge sharing hinder cross-organizational collaboration, leading to siloed expertise.   To address these challenges, we propose \textbf{AgentNet}, a decentralized, Retrieval-Augmented Generation (RAG)-based framework that enables LLM-based agents to autonomously evolve their capabilities and collaborate efficiently in a Directed Acyclic Graph (DAG)-structured network.   Unlike traditional multi-agent systems that depend on static role assignments or centralized control, AgentNet allows agents to specialize dynamically, adjust their connectivity, and route tasks without relying on predefined workflows.
AgentNet’s core design is built upon several key innovations: (1) Fully Decentralized Paradigm: Removing the central orchestrator, allowing agents to coordinate and specialize autonomously, fostering fault tolerance and emergent collective intelligence. (2) Dynamically Evolving Graph Topology: Real-time adaptation of agent connections based on task demands, ensuring scalability and resilience.
(3) Adaptive Learning for Expertise Refinement: A retrieval-based memory system that enables agents to continuously update and refine their specialized skills.
By eliminating centralized control, AgentNet enhances fault tolerance, promotes scalable specialization, and enables privacy-preserving collaboration across organizations.      Through decentralized coordination and minimal data exchange, agents can leverage diverse knowledge sources while safeguarding sensitive information.      Experimental results demonstrate that AgentNet outperforms traditional centralized multi-agent systems, significantly improving efficiency, adaptability, and scalability in dynamic environments, making it a promising foundation for next-generation autonomous, privacy-respecting multi-agent ecosystems.
\end{abstract}

% keywords can be removed

\section{Introduction}
% 1.  Challenges in existing systems:
% - Centralized control causing bottlenecks, scalability issues, and single points of failure.
% - Privacy and proprietary concerns in multi-agent collaboration.

% 2.  Proposed solution: AgentNet – A decentralized framework that:
% - Adaptive agent evolution (agents dynamically evolve and refine expertise).
% - Decentralized architecture (eliminates the reliance on a central orchestrator).
% - Dynamic task allocation (enables flexible, efficient task routing using a Directed Acyclic Graph).
% - Privacy protection (agents store and share minimal task-relevant information).

% 3.  Key innovations:
% - Fully decentralized architecture.
% - Dynamically evolving graph topology.
% - Adaptive learning via retrieval-based memory (RAG) for expertise refinement.
% - Privacy-preserving methods for decentralized collaboration.

Recently, large language models (LLMs) have demonstrated remarkable capabilities in various domains, ranging from basic text understanding to complex reasoning and multimodal integration \citep{openai2024gpt4technicalreport, touvron2023llama2openfoundation, yang2025whosmvpgametheoreticevaluation}.  Consequently, LLM-based agents have exhibited exceptional performance in numerous tasks, including scientific discovery \citep{gottweis2025aicoscientist}, automated reasoning \citep{putta2024agentqadvancedreasoning}, and website operations \citep{zhou2024tradenhancingllmagents}. However, due to the lack of collective intelligence and collaboration, LLM-based Single Agents struggle to address the complex challenges encountered in the real world. 
By leveraging collective intelligence through parallel decision-making or workflow collaboration, LLM-based Multi-Agent Systems (MAS) have emerged as a promising framework for tackling complex real-world problems \citep{Guo2024LargeLM, Sun2024LLMbasedMR, yang2024llmbasedmultiagentsystemstechniques}. However, most MAS following the workflow collaboration paradigm rely heavily on a centralized controller or a static, predefined workflow to allocate tasks among agents with fixed roles \citep{chen2023auto,hong2024metagptmetaprogrammingmultiagent,wu2023autogenenablingnextgenllm,wang2024megaagent, ye2025masgpttrainingllmsbuild}. While such designs simplify orchestration, they also introduce inherent constraints—including limited scalability, a single point of failure, and challenges to cross-organizational collaboration due to privacy and proprietary knowledge concerns.

\begin{figure*}[t]
    \centering
    \includegraphics[width=0.85\linewidth]{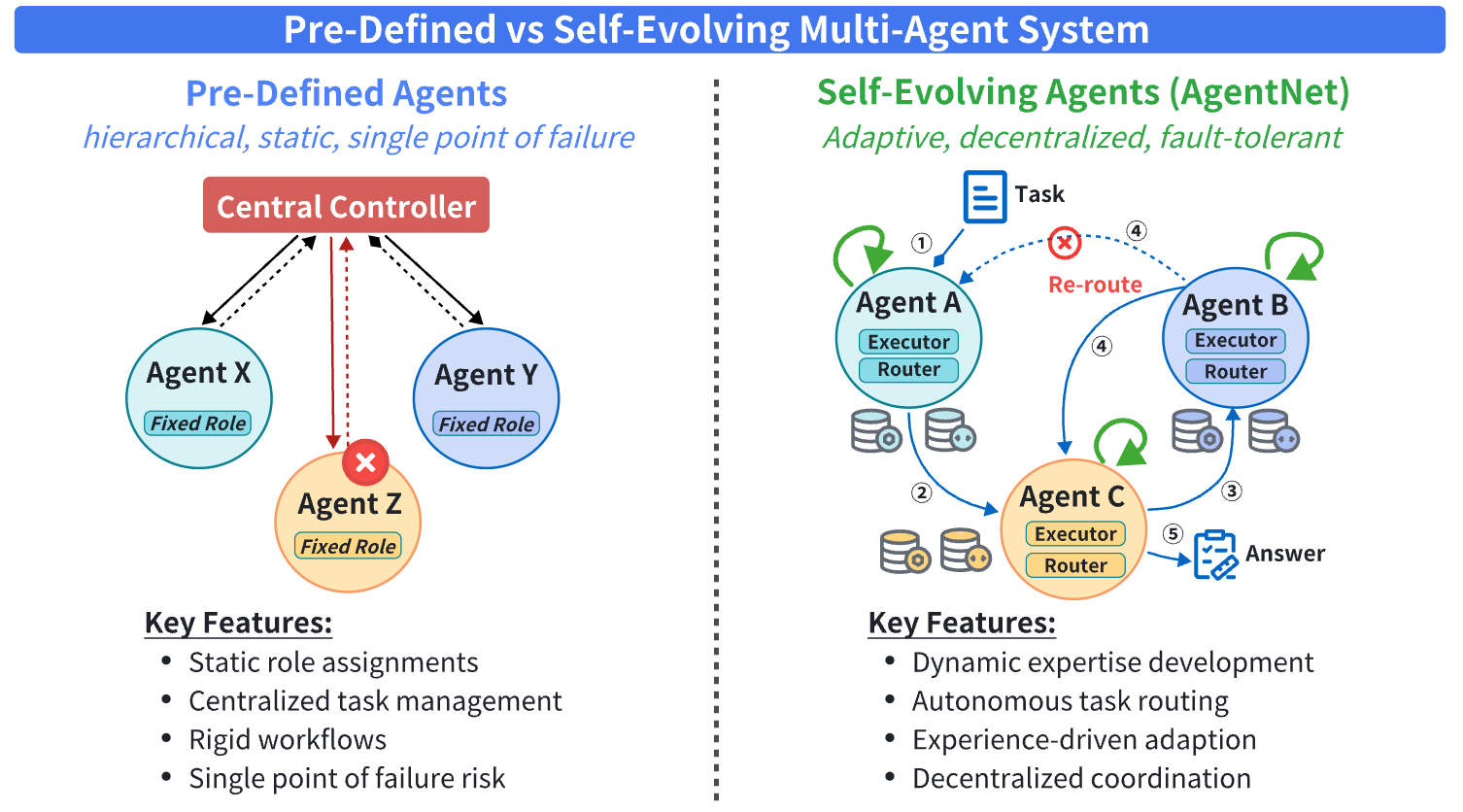}
    % \vspace{-0.2cm}
    \caption{The illustration contrasts Pre-Defined Multi-Agents (hierarchical, static, with centralized control and single point of failure) against Self-Evolving Agents/AgentNet (adaptive, decentralized, and fault-tolerant with dynamic expertise development).}
    \label{fig:compare}
    \vspace{-0.5cm}
\end{figure*}

A more critical drawback arises from the inability of these systems to adapt to real-time fluctuations in agent performance or rapidly changing task requirements. Relying on a central controller inflates deployment complexity and restricts dynamic role reassignment, rendering the system vulnerable when the controller fails or becomes overloaded. Furthermore, rigid role definitions prevent agents from flexibly leveraging their full expertise in dynamic environments, ultimately undermining both efficiency and scalability. Taken together, these limitations highlight the need for more decentralized, fault-tolerant approaches that support dynamic task allocation, enhance adaptability, and safeguard privacy across organizational boundaries.

Beyond the scalability and failure-tolerance issues previously discussed, centralized architectures become even more problematic when organizations attempt to collaborate at scale \citep{yang2024llmbasedmultiagentsystemstechniques,shi2025privacyenhancingparadigmsfederatedmultiagent}. Each institution—be it an enterprise, research lab, or government agency—typically holds proprietary expertise, sensitive data, or both. In a centralized setup, concerns over data ownership, privacy regulations, and inconsistent governance often create barriers that prevent free exchange of knowledge. As a result, LLM-based agents contributed by multiple organizations remain siloed, unable to fully capitalize on each other’s specialized capabilities or datasets. This fragmentation not only hampers collective intelligence but also highlights the urgency of developing secure, decentralized collaboration mechanisms. By enabling each participant to maintain and share only the minimal necessary information, these mechanisms address data confidentiality requirements while still allowing for a richer, more collaborative multi-agent ecosystem.

To address these challenges in multi-agent systems, we propose AgentNet, a novel framework designed to foster adaptive agent evolution, optimize task coordination, and preserve privacy. By eliminating the reliance on a central orchestrator, AgentNet enables agents to dynamically reconfigure their connections and redistribute tasks, forming a self-organizing, fault-tolerant architecture. Within this architecture, tasks are efficiently routed via a Directed Acyclic Graph (DAG) \citep{kahn1962topological,ahuja1993network}, which supports flexible collaboration and prevents cyclic dependencies.

Unlike traditional MAS frameworks that fix each agent’s role, AgentNet incorporates a retrieval-based RAG \citep{lewis2020retrieval,gao2023retrieval, zhou2024tradenhancingllmagents} memory mechanism to refine agent expertise over time. Each agent maintains a limited-capacity pool of successful task trajectories; when a new task arises, it retrieves the most relevant trajectories through few-shot learning, thus improving decision-making. To prevent memory overflow, agents autonomously prune less pertinent trajectories, ensuring the retention of valuable knowledge. This dynamic specialization strategy not only streamlines task allocation and agent adaptation but also supports a highly scalable and privacy-respecting environment for multi-agent collaboration.

AgentNet’s core design is built upon several key innovations:\vspace{-0.2cm}
\begin{itemize}[itemsep=0em]
    \item \textbf{Fully Decentralized Paradigm}: 
    By removing the need for a central orchestrator, AgentNet fosters emergent collective intelligence. Decision-making authority is distributed across all agents, thereby eliminating single points of failure and allowing each agent to coordinate, delegate, and specialize as conditions evolve. This approach leads to a self-organizing and fault-tolerant architecture that can rapidly respond to new tasks and unforeseen challenges. This decentralized setup also encourages emergent collective intelligence—in other words, agents can collectively discover and refine optimal strategies rather than waiting for instructions from a central controller.
    
    \item \textbf{Dynamically Evolving Graph Topology}: 
    AgentNet employs a network structure in which both nodes (agents) and edges (agent-to-agent connections) adapt in real time based on task demands and agent performance. Rather than relying on fixed workflows, the system continuously reconfigures its topology to optimize information flow and task distribution, ensuring scalability and resilience in complex, changing environments.
    
    \item \textbf{Adaptive Learning Mechanism for Expertise Refinement}: 
    AgentNet’s third innovation is its retrieval-based memory system, enabling agents to capture and update knowledge from successful task trajectories. This mechanism continuously refines each agent’s specialized skills without altering the network’s topology, allowing agents to avoid over-reliance on outdated information and sustain high performance in dynamic scenarios.
\end{itemize}

Moreover, each of these three innovations inherently enhances data privacy. By eliminating a central orchestrator, every agent stores and processes knowledge locally, sharing only minimal task-relevant metadata.  The dynamic graph topology further confines data flow to necessary agent-to-agent interactions, reducing the exposure of sensitive information.  Meanwhile, the retrieval-based memory mechanism restricts how much and how long data is retained, pruning outdated trajectories so that only high-value knowledge persists.  Together, these design choices safeguard privacy and intellectual property, particularly crucial for cross-organizational collaborations.

% AgentNet’s decentralized nature effectively preserves data privacy by ensuring that each agent independently stores and processes its local knowledge without relying on centralized data repositories.     This is achieved by maintaining a retrieval-based memory mechanism where agents store only relevant task trajectories locally, preventing sensitive data from being aggregated in vulnerable centralized locations.     Moreover, decision-making is decentralized, with agents communicating only necessary metadata rather than raw data, thereby minimizing the risk of information leakage.     Task coordination is managed through peer-to-peer interactions, and the DAG-based routing framework ensures that information dissemination remains localized to relevant agents.     Additionally, knowledge sharing between agents is selective, transmitting only sub-task result rather than complete proprietary information.     These mechanisms collectively enable AgentNet to provide robust privacy protection, making it well-suited for scenarios requiring cross-organizational collaboration while safeguarding sensitive data.

Our experimental evaluation shows that AgentNet significantly outperforms traditional LLM-based multi-agent frameworks in dynamic environments, demonstrating improved task efficiency, specialization stability, and adaptive learning speed. These results highlight the effectiveness of decentralized evolutionary coordination in large-scale AI ecosystems.

\section{Related Work}
\subsection{LLM-based Multi-Agent Systems}
The development of LLM-based multi-agent systems (LaMAS) \citep{yang2024llmbasedmultiagentsystemstechniques} has advanced rapidly in recent years. Early frameworks, such as AutoGen \citep{wu2023autogenenablingnextgenllm} and MetaGPT \citep{hong2024metagptmetaprogrammingmultiagent}, made significant strides in establishing foundational architectures for orchestrating multiple LLM agents through structured workflows. AutoGen provided a flexible framework for defining agent interactions, while MetaGPT incorporated software development principles to enhance collaboration. These centralized frameworks proved effective for managing multi-agent interactions. However, they also faced inherent challenges, including limited scalability, single points of failure, and difficulty in dynamically adapting to evolving tasks or incorporating new expertise.

In response to these limitations, more recent frameworks such as AgentScope \citep{gao2024agentscopeflexiblerobustmultiagent} and MegaAgent \citep{wang2024megaagent} have focused on improving robustness and scalability. AgentScope introduced modular design patterns to enhance system reliability, while MegaAgent employed hierarchical structures to scale agent interactions. Although these frameworks offer improvements, they still operate under centralized control paradigms, with a master agent delegating tasks, which continues to lead to scalability bottlenecks and single points of failure. Moreover, existing LaMAS implementations predominantly utilize single-source LLMs, lacking the integration of heterogeneous models. Their workflows are typically static, unable to dynamically allocate resources based on task complexity, further constraining adaptability.

In contrast, AgentNet introduces a novel decentralized approach, addressing these challenges by enabling agents to autonomously refine their expertise and dynamically allocate resources. AgentNet supports scalable, fault-tolerant collaboration without reliance on a central orchestrator, overcoming the limitations of centralized frameworks.

\subsection{Evolutionary Agent Systems}
Inspired by natural evolution, recent researchers have explored evolutionary approaches to automate and optimize agent behaviors and workflows in LaMAS. Existing efforts can be broadly categorized into the following areas:

\vspace{-0.1cm}
\begin{itemize}[itemsep=0em]
    \item \textbf{Prompt Evolution and Optimization} – Techniques such as PromptBreeder \citep{fernando2023promptbreeder}, DsPy \citep{khattab2023dspy} and AgentPrune \citep{zhang2024cutcrapeconomicalcommunication} apply evolutionary algorithms to iteratively refine prompt generation, improving task performance through better input design.
    \item \textbf{Inter-Agent Topology Optimization} – Systems like GPTSwarm \citep{zhuge2024gptswarm}, DyLAN \citep{liu2024dynamicllmpoweredagentnetwork}, and G-Designer \citep{zhang2024g} focus on evolving the structural organization of agent interactions. These works aim to optimize communication patterns, task allocation, and collaboration efficiency within multi-agent networks.
    \item \textbf{Agent Role and Persona Specialization} – Frameworks such as AgentVerse and MorphAgent \citep{chen2023agentverse,lu2024morphagent} refine agent roles and profiles, enabling more effective specialization and coordination among agents in complex tasks.
\end{itemize}

While these evolutionary approaches have shown promise, they primarily focus on individual agent adaptation rather than collective coordination. Additionally, they still tend to operate within centralized control structures, which limits their scalability and dynamic adaptability. Recent frameworks like AgentSquare \citep{shang2024agentsquare} and AFlow \citep{zhang2024aflow} have begun to formalize automated design processes for agentic systems, improving system-level orchestration and workflow automation. Another key direction is self-adaptive agent architectures, where agents adjust their strategies in real-time based on feedback and accumulated experience. For example, EvoMAC \citep{hu2024evomac} combines reinforcement learning with evolutionary algorithms to optimize agent decision-making and policy updates.

However, these approaches are often limited to single-agent adaptation and lack mechanisms for decentralized specialization and coordination across large-scale agent collectives. While EvoMAC and other systems focus on optimizing individual agents, they are not designed for scalable, multi-agent, decentralized collaboration. 
In contrast, AgentNet integrates evolutionary learning with decentralized control, enabling heterogeneous agents to dynamically evolve their roles, adapt their strategies in real-time, and collaborate flexibly across a large-scale multi-agent system. This integration of evolutionary learning with decentralized control makes AgentNet a more suitable framework for real-time, adaptive, and scalable multi-agent collaboration.

% In summary, while evolutionary approaches introduce promising adaptability and optimization techniques, most existing systems remain centralized, with static workflows and limited support for dynamic multi-agent collaboration. There is a critical need for decentralized architectures that enable heterogeneous agents to evolve their roles, coordinate flexibly, and collaboratively adapt workflows in real-time. AgentNet fills this gap by combining evolutionary learning with decentralized control, enabling scalable, adaptive, and fault-tolerant multi-agent collaboration.

\begin{figure*}[t]
    \centering
    \includegraphics[width=0.95\linewidth]{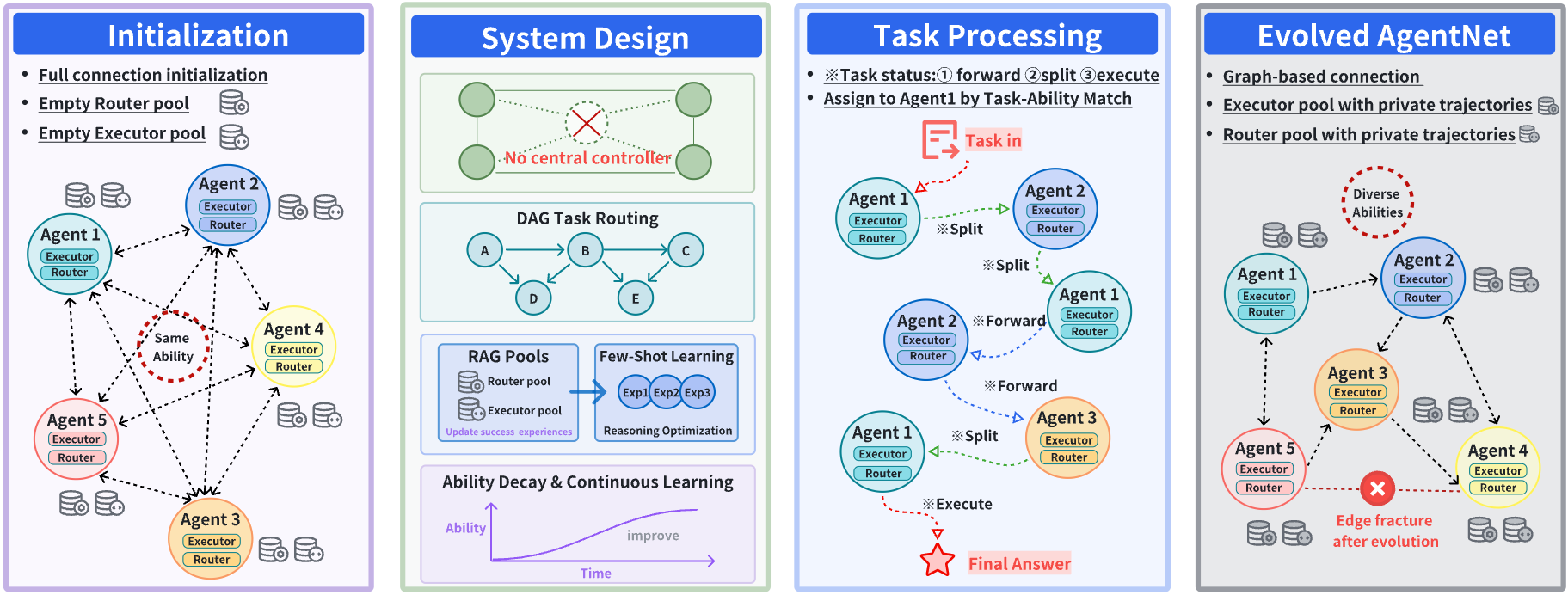}
    % \vspace{-0.2cm}
    \caption{Illutration of AgentNet. Initially, agents are fully connected and equipped with executors and routers. The system eliminates the need for a central controller, using a DAG for dynamic task routing and agents leveraging RAG pools and few-shot learning. In the evolved phase, the network adapts with agents developing private trajectories and diversified abilities, ensuring scalability, fault tolerance, and continuous evolution of expertise..}
    \label{fig:Illutration of AgentNet}
    \vspace{-0.5cm}
\end{figure*}

\section{Methodology}
\subsection{Overview of AgentNet Architecture}
Unlike traditional MAS frameworks with fixed agent roles and rigid workflows using central coordinators, AgentNet creates a privacy-preserving, collective intelligence multi-agent system with high scalability and failure-tolerance by leveraging an innovative framework, consisting of a fully decentralized network architecture, a dynamic task allocation mechanism, and an adaptive agent learning method.

We begin with a brief introduction of AgentNet, including notation and basic architectures of agents employed. 

Formally, we define AgentNet as a tuple $\mathcal{G} = (\boldsymbol{\mathcal{A}}, \boldsymbol{\mathcal{E}})$, where $\boldsymbol{\mathcal{A}} = \{a_1, a_2, ..., a_n\}$ represents the set of autonomous agents, $\boldsymbol{\mathcal{C}} = \{c_1, c_2, ..., c_n\}$ represents each agent's ability, and $\boldsymbol{\mathcal{E}} \subseteq \boldsymbol{\mathcal{A}} \times \boldsymbol{\mathcal{A}}$ represents the communication connections between agents, specifically $e_{i,j}\in \boldsymbol{\mathcal{E}}$ referring to a unidirectional connection from Agent $a_i$ to Agent $a_j$.

For each agent $a_i \in \boldsymbol{\mathcal{A}}$ contains two key components. $rou_i$ is an agent router, responsible for analyzing received routing queries and making routing decisions. $exe_i$ is an agent executor, responsible for responding to executing queries through operations and tools. 

The two components mentioned above are underpinned by a substantial LLM that leverages its extensive knowledge and understanding to solve specific problems. Furthermore, both $rou_i$ and $exe_i$ in $a_i$ maintain fixed-size memory modules $M_i^{rou}$ and $M_i^{exe}$, respectively, providing $a_i$ with powerful adaptive evolutionary capabilities by storing and utilizing the agent's experiences through the RAG mechanism. 

For optimization, AgentNet will be given a series of tasks denoted as $\boldsymbol{T}=\{t_1, t_2, ..., t_M\}$ to resolve, along with an evaluation function $Eval(\cdot)$. The optimization goal of AgentNet is to maximize the evaluated score by $Eval(\cdot)$ for the solution output by AgentNet, specifically optimizing $\boldsymbol{\mathcal{A}}$ and $\boldsymbol{\mathcal{E}}$, as the following formula:

\begin{equation}
    \mathcal{G^*} = (\boldsymbol{\mathcal{A^*}}, \boldsymbol{\mathcal{E^*}}) = \mathop{\arg\max}\limits_{\boldsymbol{\mathcal{A}}, \boldsymbol{\mathcal{E}}}\ Eval(\mathcal{G}, \boldsymbol{T}).
\end{equation}

The innovation of AgentNet emerges from the synergistic integration of three key mechanisms: (1) AgentNet realized a fully decentralized network architecture by distributing decision-making authority across all agents, leading to a high failure tolerance and privacy-preserving MAS. (2) By using a dynamic task allocation mechanism, AgentNet can optimize workload distribution based on agent capabilities and the current system state flexibly. (3) The adaptive learning in AgentNet can achieve continuous specialization of agents, making the whole MAS more scalable and adaptive. Therefore, we can create a self-organizing system capable of handling complex tasks while preserving privacy and adapting to changing environments.

\begin{algorithm}[t]
\caption{AgentNet System}
\begin{algorithmic}[1]
\Require Task set $T = \{t_1, t_2, \ldots, t_M\}$
\Ensure Optimized network $\mathcal{G}^* = (\boldsymbol{\mathcal{A}}^*, \boldsymbol{\mathcal{E}}^*)$
\State Initialize $\boldsymbol{\mathcal{A}} = \{a_1, a_2, \dots, a_n\}$, $\boldsymbol{\mathcal{E}}$, $\boldsymbol{\mathcal{C}} = \{c_1, c_2, \dots, c_n\}$, $w_0$, $M_i^{rou}$ and $M_i^{exe}$ $\forall a_i \in \boldsymbol{\mathcal{A}}$
\For{each task $t_{m+1} \in T$}
    \State \textcolor{blue}{// 1. Task allocation and processing}
    \State $c_{t_{m+1}} \gets \Phi(o_{t_{m+1}})$, $a_{curr} \gets \arg\max_{a_i \in \mathcal{A}_m} \text{sim}(c_{t_{m+1}}, c_i^m)$
    \State $task\_state \gets (o_{t_{m+1}}, \emptyset, p_{t_{m+1}})$, $visited \gets \emptyset$, $finished \gets false$
    \While{not $finished$ and $a_{curr} \notin visited$}
        \State $visited \gets visited \cup \{a_{curr}\}$
        \State $fragments^{rou} \gets \text{Select}(M_{curr}^{rou}, t_{m+1}, k)$
        \State $action \gets \mathcal{F}_{act}(o_{t_{m+1}}, c_{t_{m+1}}, \mathcal{F}_{reason}(o_{t_{m+1}}, c_{t_{m+1}}, fragments^{rou}), fragments^{rou})$
        
        \If{$action = \mathcal{O}_{fwd}$}
            \State $a_{curr} \gets \arg\max_{a_k \in \mathcal{A}_m \setminus \{a_{curr}\}} \text{sim}(c_{t_{m+1}}, c_k^m)$            
        \ElsIf{$action = \mathcal{O}_{split}$}
            \State $subtasks \gets \text{DecomposeTask}(t_{m+1})$
            \State $task\_state.context \gets task\_state.context \oplus \text{ProcessSubtasks}(subtasks, a_{curr}, \mathcal{A}_m)$
            \State $finished \gets \text{AllSubtasksCompleted}(subtasks)$            
        \Else
            \State $task\_state.context \gets task\_state.context \oplus \text{ExecuteTask}(a_{curr}, t_{m+1}, \text{Select}(M_{curr}^{exe}, t_{m+1}, k))$
            \State $finished \gets true$
        \EndIf
    \EndWhile
    
    \State \textcolor{blue}{// 2. Network update}
    \For{each interacting pair $(a_i, a_j)$}
        \State $w_{m+1}(i, j) \gets \alpha \cdot w_m(i, j) + (1-\alpha) \cdot S(a_i^{m+1}, a_j^{m+1}, t_{m+1})$
    \EndFor    
    \State $\boldsymbol{\mathcal{E}}_{m+1} \gets \{(a_i^{m+1}, a_j^{m+1}) \mid w_{m+1}(i, j) > \theta_w\}$
    \State \textcolor{blue}{// 3. Agent capability and memory update}
    \For{each participating agent $a_i$}
        \State $c_i^{m+1} \gets \beta \cdot c_i^m + (1-\beta) \cdot \Delta c_i^{m+1}$, Update $M_i^{rou}$ and $M_i^{exe}$
    \EndFor
\EndFor
\State \Return $\mathcal{G}^* = (\boldsymbol{\mathcal{A}}, \boldsymbol{\mathcal{E}})$
\end{algorithmic}
\end{algorithm}

\subsection{Decentralized Network Topology}

\begin{figure}[h]
    \vspace{-0.2cm} 
    \begin{minipage}{0.5\textwidth}
    As illustrated in Figure ~\ref{fig:Dual-role agent architecture.}, AgentNet employs a dual-role design. Each agent $a_i$ is equipped with a router $rou_i$ to facilitate routing decisions and an executor $exe_i$ to execute specific tasks. We will introduce details of the router and executor in the following sections. In essence, the router within each agent endows AgentNet with a fully decentralized network structure because the routing decisions are made independently by each agent, without relying on a central authority or coordinator. 
    \vspace{+0.3cm} 
    
    This contrasts with traditional LLM based Multi-Agent Systems that typically depend on a centralized controller to manage the coordination and allocation of tasks. Each agent in AgentNet autonomously determines how to route tasks to other agents based on its local knowledge and the task requirements, ensuring that decision-making is distributed across the network and that there is no single point of control, thus achieving full decentralization.

    \end{minipage}
    \hfill
    \begin{minipage}{0.5\textwidth}
        \centering
        \includegraphics[width=0.95\linewidth]{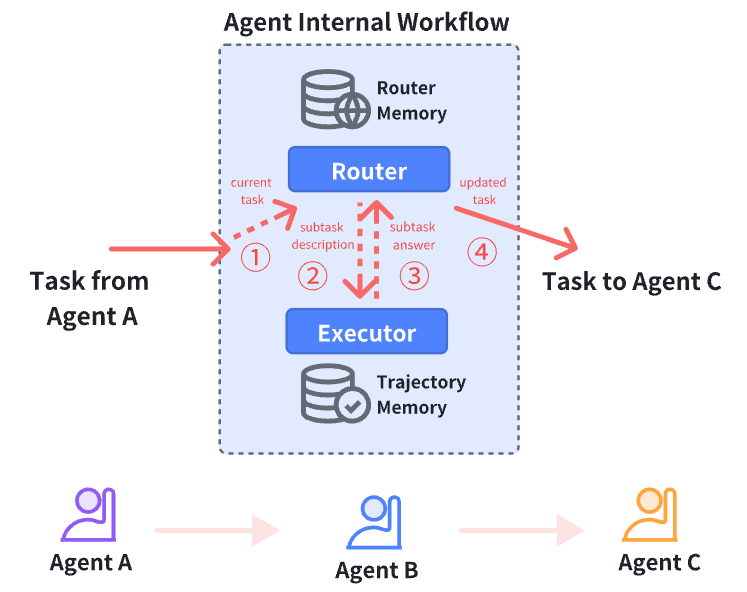}
        \caption{Dual-role agent architecture.}
        \label{fig:Dual-role agent architecture.}
    \end{minipage}
\end{figure}
\vspace{-0.1cm}

Mathematically, we represent the architecture of AgentNet as $\mathcal{G}_m = (\boldsymbol{\mathcal{A}_m}, \boldsymbol{\mathcal{E}_m})$ when given the $m+1$-th task $t_{m+1}$ after completing task $t_{m}$, where $\boldsymbol{\mathcal{A}_m} = \{a_{1}^{m}, a_{2}^{m}, ..., a_{n}^{m}\}$ represents the states of agents after task $t_{m}$ and $\boldsymbol{\mathcal{E}}_m \subseteq \boldsymbol{\mathcal{A}_m} \times \boldsymbol{\mathcal{A}_m}$ represents the set of directed edges between agents and each edge $e_{i,j}^{m}$ means a directed edge from $a_{i}^{m}$ to $a_{j}^{m}$. A weight matrix $w_{m}$ will be maintained throughout all the tasks before $t_{m+1}$ to weight the connection between agents, namely $w_{m}(i, j)$. After completing $t_{m+1}$, $w_{m+1}$ can be updated using the following formula from $w_{m}$:
\begin{equation}
w_{m+1}(i, j) = \alpha \cdot w_{m}(i, j) + (1-\alpha) \cdot  S(a_i^{m+1}, a_j^{m+1}, t_{m+1}),
\end{equation}

where $\alpha \in [0,1]$ is a decay factor that balances historical performance with recent interactions, and $S(a_i^{m+1}, a_j^{m+1}, t_{m+1})$ is a success metric for task $t_{m+1}$ routed from agent $a_i^{m+1}$ to $a_j^{m+1}$. This adaptive weighting mechanism ensures that the network continuously refines its structure based on operational experience.

Over tasks, the weight matrix $w_m$ will evolve based on collaborative success, and the edges with a lower weight than a hyper-parameter threshold $\theta_w$ are periodically pruned:
\begin{equation}
\boldsymbol{\mathcal{E}_{m+1}} = \{(a_i^{m+1}, a_j^{m+1}) \mid w_{m+1}(i, j)> \theta_w\}.
\end{equation}

This pruning mechanism ensures that the network maintains efficient pathways while eliminating unproductive connections, optimizing both communication overhead and routing efficiency.

\subsection{Adaptive Learning and Specialization}
AgentNet's adaptive learning mechanism facilitates continuous improvement and specialization of agents based on their task experiences, without the need for explicit role assignment. This process enables agents to gradually develop expertise in specific domains, differentiating AgentNet from static multi-agent systems and allowing it to adapt to evolving requirements over time.

Agents in AgentNet follow the \textit{ReAct} (Reasoning + Acting) framework \cite{yao2023reactsynergizingreasoningacting, zhou2024tradenhancingllmagents}, which empowers agents to reason about a given query and its context before deciding appropriate actions for the executor modules. In addition to the given query and its context, the agent also retrieves relevant trajectory fragments from its memory modules to enhance reasoning and acing. The retrieval process is performed using a Retrieval-Augmented Generation (RAG) mechanism \citep{lewis2020retrieval,gao2023retrieval, zhou2024tradenhancingllmagents}, which allows the agent to leverage past experiences to generate informed decisions and actions for new tasks.

In the AgentNet, Each agent $a_i\in \boldsymbol{\mathcal{A}}$ maintains two memory modules $M_i^{rou}$ and $M_i^{exe}$ for its router module $rou_i$ and $exe_i$, which store local trajectory fragments from prior tasks corresponding to the specific steps where $a_i$ was actively involved instead of storing the whole task trajectories cooperated by all agents.

Formally, each entry in $M_i^{rou}$ and $M_i^{exe}$ is the local step fragment represented as: $f^r = (o^r, c^r, a^r)$, where $r$ represents this entry belongs to $rou_i$ (when $r=rou$) or $exe_i$ (when $r=exe$). $o^r$ denotes the observation, namely the query of the corresponding task, and $c^r$ represents the context of the corresponding task so far (i.e., partial trajectory before this step), and $a^r$ is the action or response of the agent. These step fragments are collected from different tasks in which the agent has participated and serve as experiential knowledge for future reasoning.

When agent $a_i^m$ receives a new task $t_{m+1}$ to solve, it retrieves the $k$ most relevant fragments from both memory modules. For each module type $r \in \{rou, exe\}$, the retrieval process is defined as:

\begin{equation}
\begin{split}
\text{Select}(M_i^r, t_{m+1}, k) = \{f_1^r, f_2^r, \dots, f_k^r\} \subset M_i^r, \text{ such that } \\
\forall f_j^r \in \text{Select}(M_i^r, t_{m+1}, k) \text{ and } \forall f \in M_i^r \setminus \text{Select}(M_i^r, t_{m+1}, k): \\
\text{sim}(\text{embed}(o_{f_j^r}^r, c_{f_j^r}^r), \text{embed}(o_{t_{m+1}}^r, c_{t_{m+1}}^r)) \geq \text{sim}(\text{embed}(o_f^r, c_f^r), \text{embed}(o_{t_{m+1}}^r, c_{t_{m+1}}^r)).
\end{split}
\end{equation}

Here, \(\text{embed}(\cdot)\) is a semantic embedding function that projects the input context into a high-dimensional vector space, and the fragments with the highest relevance are retrieved to inform the agent's reasoning or action for both routing and execution processes.

Both the reasoning and acting processes are enhanced by the retrieval of historical task fragments, allowing the agent to make better decisions based on prior experiences.
The reasoning function for each module type is modeled as:

\begin{equation}
\mathcal{R}_{a_i}(t_{m+1}, r) = \mathcal{F}_{\text{reason}}(o_{t_{m+1}}, \text{c}_{t_{m+1}}, \{f_j^r\}_{j=1}^k),
\end{equation}

where $\mathcal{F}_{\text{reason}}$ represents the large language model that serves as the backbone of the LLM Agent, processing the inputs to generate reasoned decisions.
The reasoning function $\mathcal{R}_{a_i}$ takes the current observation and question \(o_{t_{m+1}}\), the historical context \(\text{c}_{t_{m+1}}\) representing the partial task trajectory and interactions up to the current point, and the retrieved fragments \(\{f_j^r\}_{j=1}^k\) as input to generate the reasoning output. The fragments allow the agent to reason based on prior experiences that are most relevant to the current situation.

Once the reasoning process has been completed, the agent executes the chosen action. The action is informed by the reasoning output, which can be expressed as:

\begin{equation}
\mathcal{A}_{a_i}(t_{m+1}, r) = \mathcal{F}_{\text{act}}(o_{t_{m+1}}, \text{c}_{t_{m+1}}, \mathcal{R}_{a_i}(t_{m+1}, r), \{f_j^r\}_{j=1}^k),
\end{equation}

where $\mathcal{F}_{\text{act}}$ represents the large language model that serves as the backbone of the LLM Agent, translating reasoning into concrete operations.
The $\mathcal{A}_{a_i}(t_{m+1}, r)$ function utilizes the reasoning output $\mathcal{R}_{a_i}(t_{m+1}, r)$ along with the retrieved memory fragments to determine the appropriate action. The specific action depends on the module type: for $r=rou$, the router module may produce actions such as forwarding the task to another agent or splitting it into subtasks; for $r=exe$, the executor module generates a single-step operation or response to directly address the final answer.

To manage memory effectively, each agent employs a dynamic memory management strategy. As the agent receives new tasks, it evaluates the trajectories stored in its memory modules and decides which to retain or prune. This evaluation is based on reasoning about the task context, historical usage patterns, and the relevance of each trajectory to future tasks. Several factors influence the decision-making process, including the frequency of use, recency of tasks, and the uniqueness of the trajectory. The agent assesses these factors through a prompt-based reasoning process that helps determine the utility of each trajectory. When a memory module reaches its capacity limit \(C_{\text{max}}\), the agent compares the new trajectory with the existing ones in that memory module and selects the least useful trajectory to remove, thus ensuring the memory pool remains focused on high-quality and relevant knowledge.

Through this adaptive and memory-driven learning process, agents in AgentNet continuously refine their expertise and specialize in the areas where they excel. This specialization occurs naturally over time, allowing the system to self-organize and adapt to the demands of a wide variety of tasks.

\subsection{Dynamic Task Allocation}
\begin{figure*}[t]
    \centering
    \includegraphics[width=0.95\linewidth]{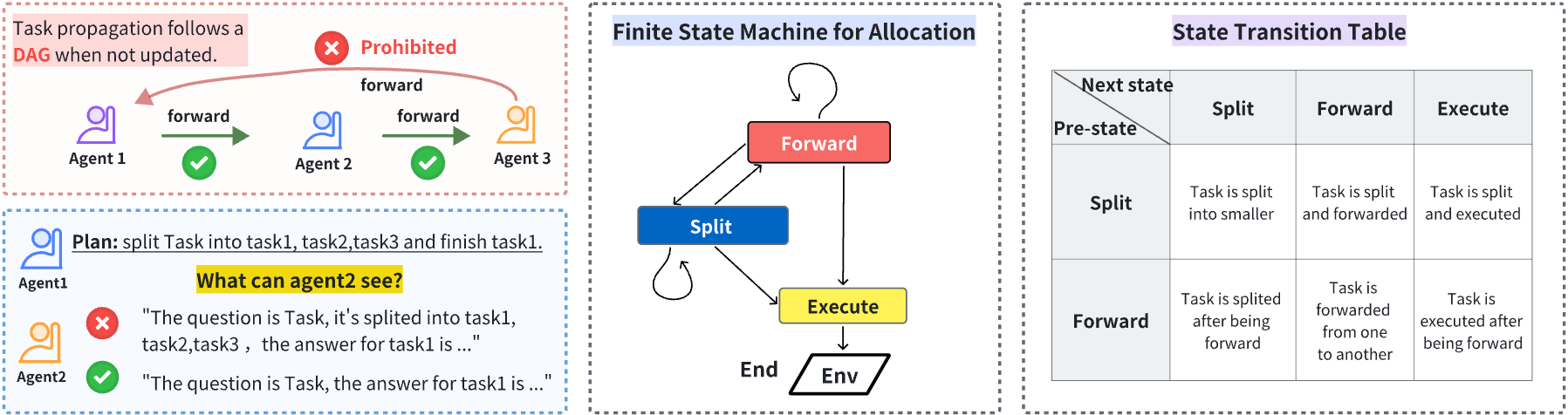}
    % \vspace{-0.2cm}
    \caption{Details of Dynamic Task Allocation.}
    \label{fig:Dynamic Task Allocation}
    \vspace{-0.3cm}
\end{figure*}

The dynamic task allocation mechanism in AgentNet enables efficient distribution of tasks without centralized coordination, creating a responsive system that optimizes both performance and load balancing. This decentralized approach to task routing represents a significant advancement over static assignment strategies employed in traditional multi-agent frameworks.

Each task $t \in T$ is formally represented as a tuple $t = (o_t, c_t, p_t)$, where $o_t$ contains the task description in natural language, $c_t$ is a vector of capability requirements, and $p_t$ denotes the priority level. To efficiently process a new task $t_{m+1}$ after completing task $t_m$, AgentNet employs a sophisticated mechanism to select the most suitable initial agent.

Agent capability representation and matching form the foundation of task allocation. Each agent $a_i^m$, after completing task $t_m$, possesses a capability vector $c_i^m$ that is dynamically updated through task performance during system operation. In the initial allocation phase, the system selects an entry agent for $t_{m+1}$ using the following formula:
\begin{equation}
a_{initial} = \underset{a_i \in \mathcal{A}_m}{\text{argmax}} \{ \text{sim}(c_{t_{m+1}}, c_i^m) \},
\end{equation}
where $c_{t_{m+1}} = \Phi(o_{t_{m+1}})$ represents the capability requirements of task $t_{m+1}$, $c_i^m$ denotes the capability vector of agent $a_i$, and $\text{sim}(\cdot, \cdot)$ is a similarity function measuring the match between task requirements and agent capabilities. The capability requirements are determined through different methodologies depending on task complexity:

\begin{equation}
c_{t_{m+1}} = \begin{cases}
\Phi_{atomic}(t_{m+1}), & \text{for atomic tasks} \\
\Phi_{compound}(t_{m+1}), & \text{for compound tasks}.
\end{cases}
\end{equation}

For atomic tasks, the system employs function $\Phi_{atomic}$ that maps task properties to capability requirements based on predefined heuristics. For compound tasks, function $\Phi_{compound}$ leverages an instruction set containing carefully crafted prompts that guide the large language model in analyzing task descriptions and inferring the required capability vectors. The system subsequently ranks all agents according to capability matching scores and selects the highest-scoring agent as the initial agent.

Once a task is assigned to the initial agent, this agent determines how to process the task based on the reasoning results from its router module $rou_i$. As illustrated in Figure \ref{fig:Dynamic Task Allocation}, the agent can perform three operations:

\begin{enumerate}
    \item \textbf{Forward} ($\mathcal{O}_{\text{fwd}}$): Transfer the task unchanged to another more suitable agent, maintaining the task's original state and preserving the Directed Acyclic Graph (DAG) property of the routing path. Forwarding decisions are based on analyzing the gap between the current agent's capabilities and the task requirements, as well as evaluating the capability vectors of other agents in the network.
    
    \item \textbf{Split} ($\mathcal{O}_{\text{split}}$): Decompose the task into subtasks, execute portions matching the agent's expertise, and route the remaining subtasks to an appropriate agents. Subtask routing follows this formula:
    \begin{equation}
    a_{next} = \underset{a_k \in \mathcal{A}_m \setminus \{a_i\}}{\text{argmax}} \{ \text{sim}(\Phi(o_{t_{m+1}}), c_k^m) \},
    \end{equation}
    where $\Phi(o_{t_{m+1}})$ represents the capability requirements derived from the observation of subtask $j$, determined through the current agent's task decomposition reasoning, and $\mathcal{A}_m \setminus \{a_i\}$ denotes the set of all agents excluding the current one.
    
    \item \textbf{Execute} ($\mathcal{O}_{\text{exec}}$): Complete the entire task without further delegation.
\end{enumerate}

A key design feature in the system is that when an agent chooses to split a task, it only forwards the results of the subtasks it has completed, and not the reasoning behind the decomposition. This prevents the transfer of unnecessary information and ensures that task decomposition errors made by one agent do not propagate to other agents in the network.

The agent capability vector $c_i^m$ is updated based on task execution history and success rates, using the following formula:
\begin{equation}
c_i^{m+1} = \beta \cdot c_i^m + (1-\beta) \cdot \Delta c_i^{m+1},
\end{equation}

where $\beta \in [0,1]$ is a decay factor balancing historical capabilities with newly acquired ones, and $\Delta c_i^{m+1}$ represents the new capability contribution demonstrated by the agent in task $t_{m+1}$, calculated by analyzing the types of operations successfully executed by the agent and the quality of results.

Furthermore, the task's state is updated only when an agent completes a part of the task (whether by executing or splitting it). When the agent completes a subtask, it updates the context and forwards it to the next agent:
\begin{equation}
\text{context}_{\text{updated}} = \text{context}_{\text{original}} \oplus \text{result}(a_j, t_i).
\end{equation}

While the task is only being forwarded from one agent to another, its state remains unchanged, preserving the Directed Acyclic Graph (DAG) structure of the task routing path. This ensures that the task's progression avoids being trapped in an infinite loop during task forwarding, maintaining a consistent and effective routing process across different agents.

Through this dynamic task allocation mechanism, AgentNet can adaptively optimize task flow based on task characteristics and changes in agent capabilities, achieving improved overall system performance and efficient resource utilization.

\section{Experiment}

\subsection{Experimental Setup}
\paragraph{Tasks and Benchmarks}
We evaluate methods using several benchmarks across three task categories, along with custom constructed training and test sets for each benchmark:\vspace{-0.2cm}
\begin{itemize}
\item \textbf{Mathematics}: This task involves mathematical problem and is evaluated using MATH \cite{hendrycksmath2021}, which includes problems with 7 different types. The training set consists of 100 examples per type (total of 700 problems), while the test set consists of 20 examples per type (total of 140 problems). 
% The problems span five levels of difficulty, with complexity calculated based on the given complexity in the dataset and the current task's information length.

\item \textbf{Logical Question Answering}: This task tests reasoning and logical question answering abilities using the BBH (Big-Bench Hard) benchmark \cite{suzgun2022challengingbigbenchtaskschainofthought}. The training set follows the MorphAgent setup, selecting 627 examples from 20 tasks. For testing, each task has 5 examples of varying difficulty, totaling 100 test problems.

\item \textbf{Function-Calling}: This benchmark evaluates the agent's ability to perform tool-augmented task planning and API usage, based on the API-Bank dataset \cite{li2023apibankcomprehensivebenchmarktoolaugmented}. We construct a training set of 100 tasks and a test set of 100 tasks, randomly sampled from the full API-Bank corpus. Since the original dataset does not include category labels, we annotate each task using GPT-4o-mini to assign one of the seven task types: \textit{health}, \textit{account}, \textit{schedule}, \textit{information}, \textit{housework}, \textit{finance}, and \textit{others}. Each task is further categorized into one of three difficulty levels, determined by prompt complexity and required toolchain length.

\end{itemize}

% \paragraph{Metrics} A range of evaluation metrics have been adopted for different tasks. For the Mathematics and the Logical Question Answering tasks, the accuracy metric is utilized to evaluate the consistency of the output answer with the true answer within the specified format. For the Coding task, the average test case pass rate (i.e., the ratio of the number of passed test cases to the total number of test cases) and the ratio of problems passed across all test cases have been employed as the evaluation metrics.

\paragraph{Baselines}
We compare AgentNet with two categories of baselines: single-agent and multi-agent frameworks:\vspace{-0.2cm}
\begin{itemize}
    \item \textbf{Single-agent frameworks:}
    These methods involve a single agent solving tasks independently without collaboration or coordination with other agents.
    \begin{itemize}
        \item \textbf{Direct}: A baseline approach where the LLM directly generates outputs.
        \item \textbf{React}: A prompting technique that elicits step-by-step reasoning from language models \cite{wei2023chainofthoughtpromptingelicitsreasoning, yao2023reactsynergizingreasoningacting}.
        \item \textbf{Synapse}: A trajectory-as-exemplar prompting method, which prompts the LLM with complete trajectories of the abstracted states and actions to improve multi-step decision-making. \cite{zheng2024synapsetrajectoryasexemplarpromptingmemory}
        \item \textbf{Self-Consistency}: A decoding strategy that samples multiple reasoning paths and selects the most consistent answer through majority voting, enhancing reliability \cite{wang2023selfconsistencyimproveschainthought}.
        \item \textbf{Self-Refinement}: An iterative approach where models critically evaluate and improve their own solutions over multiple passes, progressively enhancing solution quality \cite{madaan2023selfrefineiterativerefinementselffeedback}.
    \end{itemize}
    \item \textbf{Multi-agent frameworks:}
    These methods involve multiple agents working collaboratively to solve tasks, each contributing to different aspects of the task-solving process.

    \begin{itemize}
    \item \textbf{MorphAgent}: A framework featuring self-
    evolving agent profiles that dynamically optimize individual expertise in the profile through three metrics \cite{lu2024morphagent}.
    \item \textbf{MetaGPT}: A software development framework where specialized agents (like product manager, architect, engineer) collaborate in a waterfall workflow to complete complex engineering tasks \cite{hong2024metagptmetaprogrammingmultiagent}.
    \item \textbf{AFLOW}: A framework that optimizes agent workflows using Monte Carlo Tree Search over code-represented workflows with execution feedback \cite{zhang2024aflow}.
    \item \textbf{GPTSwarm}: A framework modeling agents as computational graphs with automatic optimization of both prompts and agent collaboration patterns \cite{zhuge2024gptswarm}.
\end{itemize}
\end{itemize}

\paragraph{Parameter Configuration}
In our implementation, we configure the LLM API with a temperature of 0.0, a maximum token limit of 2048, and a top-p value of 1.0, ensuring consistent results throughout our experiments and enabling reliable comparisons and analysis. For the memory pool experiment, we utilize the "BAAI/bge-large-en-v1.5" model to compute the similarity between task queries and database trajectories.

\subsection{Main Results}
% Define color palette
\definecolor{lightblue}{RGB}{230, 240, 255}
\definecolor{headerblue}{RGB}{180,210,255}
\definecolor{rowshade}{RGB}{240,245,255}

\begin{table}[t]
  \centering
  \scriptsize \caption{Performance comparison of different methods across various tasks. In all multi-agent methods, we set \textbf{3 agents} for each method to ensure a fair comparison. The best result is in bold, while the second is underlined.}
  % \vspace{+0.2cm}
  \label{tab:performance_comparison}
  \renewcommand{\arraystretch}{0.7}
  \scriptsize 
  \resizebox{\textwidth}{!}{
  \begin{tabular}{@{}ccc|c|c|c@{}}
    \toprule
    \midrule
    \rowcolor{headerblue}
    \textbf{Backbone} & \textbf{Category} & \textbf{Method} & \textbf{MATH(Acc/\%)} & \textbf{BBH(Acc/\%)}  & \textbf{API-Bank(Acc/\%)}\\

    \midrule
    % Single Methods
    \multirow{10}{*}{DeepSeek-V3} & \multirow{5}{*}{Single Agent} 
    & Direct & 47.86  & 69.00 & 26.00\\
    & & React & 77.14 & 88.00 & 29.00\\
    & & Synapse & 89.28 & 92.00 & 28.00\\
    & & Self-Consistency & 88.00 & 85.00 & 29.00\\
    & & Self-Refinement & 87.14 & 84.00 & 25.00\\
    \cmidrule(lr){2-6}
    
    % Multi-Agent Systems
    & \multirow{4}{*}{Multi-Agent} 
    & MorghAgent & 39.29 & 56.00 & 16.00 \\
    & & MetaGPT & 92.14 & 64.00 & 22.00\\
    & & AFLOW & \underline{91.67} & 88.00 & 28.00 \\
    & & GPTSwarm & 72.14 & 90.00 & 21.00 \\
    \rowcolor{rowshade}

    & & \textbf{AgentNet} & \textbf{92.86} & \textbf{94.00} & \textbf{30.00} \\
    
    \midrule
    % Single Methods
    \multirow{10}{*}{GPT-4o-mini} & \multirow{5}{*}{Single Agent} 
    & Direct & 31.43 & 59.00 & 15.00\\
    & & React & 55.71 & 80.00 & 24.00\\
    & & Synapse & 77.14 & 79.00 & 22.00\\
    & & Self-Consistency & 54.28 & 85.00 & 22.00\\
    & & Self-Refinement & 68.57 & 81.00 & 23.00\\
    \cmidrule(lr){2-6}
    
    % Multi-Agent Systems
    & \multirow{4}{*}{Multi-Agent} 
    & MorghAgent & 80.71 & 56.00 & 16.00\\
    & & MetaGPT & 73.57 & 53.00 & 19.00\\
    & & AFLOW & \textbf{85.00} & 75.00 & 21.00\\
    & & GPTSwarm & \textbf{85.00} & \textbf{86.00} & 13.00\\
    \rowcolor{rowshade}
    
    & & \textbf{AgentNet} &\textbf{85.00} & \textbf{86.00} & \textbf{29.00}\\

    \midrule
    % Single Methods
    \multirow{10}{*}{Qwen-turbo} & \multirow{5}{*}{Single Agent} 
    & Direct & 37.85 & 57.00 & 27.00\\
    & & React & 53.57 & 69.00 & 23.00\\
    & & Synapse & 67.14 & 68.00 & 24.00\\
    & & Self-Consistency & 64.28 & 70.00 & 28.00\\
    & & Self-Refinement & 76.43 & 74.00 & 23.00\\
    \cmidrule(lr){2-6}
    
    % Multi-Agent Systems
    & \multirow{4}{*}{Multi-Agent} 
    & MorghAgent & 16.43 & 56.00 & 9.00\\
    & & MetaGPT & 63.57 & 51.00 & 20.00\\
    & & AFLOW & \textbf{82.14} & 57.00  & 22.00\\
    & & GPTSwarm & 79.29 & \underline{75.00} & 30.00\\
    \rowcolor{rowshade}

    & & \textbf{AgentNet} & \underline{81.43} & \textbf{92.00} & \textbf{32.00}\\
    \midrule
    \bottomrule
  \end{tabular}
  }
  \vspace{-0.5cm}
\end{table}

Table~\ref{tab:performance_comparison} summarizes performance across Math, Logical QA tasks and API-Calling tasks. For Math, Logical QAs  and API-Calling tasks, accuracy is reported.
Compared to single-agent methods (e.g., Synapse, ReAct), AgentNet achieves competitive or superior performance across all tasks. While ReAct performs well on Math and Logic, its static prompting strategy limits generalization to more complex tasks.
Against multi-agent baselines, AgentNet consistently outperforms centralized frameworks such as MetaGPT, which suffers from limited scalability—e.g., only 53.00\% accuracy on Logical QA. 
% MorphAgent underperforms on Coding tasks, as it generates self-constructed test cases during training, resulting in invalid or uncompilable outputs.
AgentNet's decentralized coordination and retrieval-augmented memory contribute to its robustness across domains, particularly in tasks requiring contextual understanding and adaptive role specialization.

\subsection{Experiments on Heterogeneous Agents}
% \begin{itemize}
%   \item \textbf{Fully Homogeneous:} All agents are identical in both model and capabilities.
%   \item \textbf{LLM Heterogeneity:} Agents use different LLMs (e.g., GPT, Deepseek, Qwen), but other configurations are identical.
%   \item \textbf{Skill Heterogeneity:} Agents are initialized with different capabilities in various dimensions (e.g., some with high skills 0.6, others low 0.4), resulting in skill diversity.
%   \item \textbf{Both Heterogeneity:} A potential combination of both LLM and skill heterogeneity.
% \end{itemize}
To investigate the impact of agent diversity on performance, we designed a heterogeneity experiment across different settings on the BBH task. Agents were tested under four configurations: fully homogeneous (identical models and capabilities), LLM heterogeneity (different language models, same capabilities), skill heterogeneity (same model but varied capabilities), and a combination of both. This design allows us to isolate and analyze how model-level and capability-level diversity influence multi-agent collaboration.
\vspace{-0.2cm}

\begin{table}[h]
\centering
\caption{BBH Accuracy under Different Heterogeneity Settings (Acc\%)}
\vspace{+0.1cm}
\label{tab:heterogeneity}
\renewcommand{\arraystretch}{1.3}
% \scriptsize
\rowcolors{2}{gray!10}{white}
\begin{tabular}{c|c|c|c|c}
\hline
\rowcolor{lightgray}
\textbf{Setting} & \textbf{Fully Homogeneous} & \textbf{Skill Hetero.} & \textbf{LLM Hetero.} & \textbf{Both Hetero.} \\
\hline
3 Agents & 0.86 & 0.84 & 0.81 & 0.81\\
\hline
5 Agents & 0.79 & 0.86 & 0.85 & 0.85 \\
\hline
\end{tabular}
% \vspace{-0.2cm}
\end{table}

The results show that the impact of heterogeneity on multi-agent performance depends on team size. With 3 agents, the fully homogeneous setting performs best, while introducing either model or skill diversity reduces accuracy, suggesting uniform reasoning is more effective in small teams. However, with 5 agents, heterogeneous configurations outperform the homogeneous one, indicating that diversity enhances collaboration and complementary reasoning in larger teams. Overall, heterogeneity may introduce coordination overhead in small groups but offers clear benefits at larger scales.

\subsection{Ablation Study}
\paragraph{Router Effectiveness in AgentNet}
% To assess the utility of the AgentNet decentralized router design, the experiment was carried out with different ablation strategies and the AgentNet system. The router of each agent is responsible for two key functions: external routing, which determines which agent the task should be passed to (next agent ID), and internal routing, which decides whether an agent should execute a task, forward it, or split it into subtasks (Operations: Forward, Split, Execute).      The experiment was tested under various configurations, including "Totally Random", "Random Operations", "Random Next Agent ID", which determine how different types of routing affect the system's performance, and "Global Router", which maintains a global centralized router for task allocation. Performance was evaluated using the BBH task, with two phases: training (627ps) and testing (100ps).

% \begin{table}[h]
% \centering
% \caption{AgentNet's Router Performance on the BBH (Backbone: gpt-4o-mini)}
% % \vspace{+0.2cm}
% \setlength{\tabcolsep}{10pt}  
% \renewcommand{\arraystretch}{1.6}
% \resizebox{\textwidth}{!}{
% \begin{tabular}{c|ccccc}
% \hline
% \rowcolor{lightblue} \textbf{BBH} & \textbf{Totally Random} & \textbf{Random Operations} & \textbf{Random Next Agent ID} & \textbf{Global Router} & \textbf{AgentNet} \\
% \hline  \textbf{Train(627ps)} & 54.86 & 69.00 & 81.49 & 81.18 & \textbf{82.14} \\
% \hline
%  \textbf{Test(100ps)} & 55.00 & 71.00 & 82.00 & 83.00 & \textbf{86.00} \\
% \hline
% \end{tabular}}
% % \vspace{+0.2cm}
% \label{tab:bbh_performance}
% \vspace{-0.2cm}
% \end{table}

To evaluate AgentNet’s decentralized router, experiments were conducted comparing AgentNet with ablation configurations: "Totally Random", "Random Operations", "Random Next Agent ID" and a centralized "Global Router". Each router manages external routing (selecting the next agent) and internal routing (deciding to forward, split, or execute). 
\vspace{-0.6cm}

\begin{figure}[h]
\centering
\begin{minipage}[h]{0.48\textwidth}
\vspace{0.1cm}
Performance was tested on the BBH task (training: 627 problems, testing: 100 problems), with results in Figure~\ref{fig:router}.
AgentNet outperforms randomized methods, achieving 82.14\% accuracy during training and 86.00\% during testing. 
Randomizing operations (forward/split/execute) affects task execution more directly, randomizing next agent ID primarily results in suboptimal task delegation but does not disrupt task completion as severely. 
These results underscore the critical role of effective routing and suggest that optimizing routing decisions can significantly enhance multi-agent system performance.
\end{minipage}
\hfill
\begin{minipage}[h]{0.45\textwidth}
\centering
\includegraphics[width=0.96\linewidth]{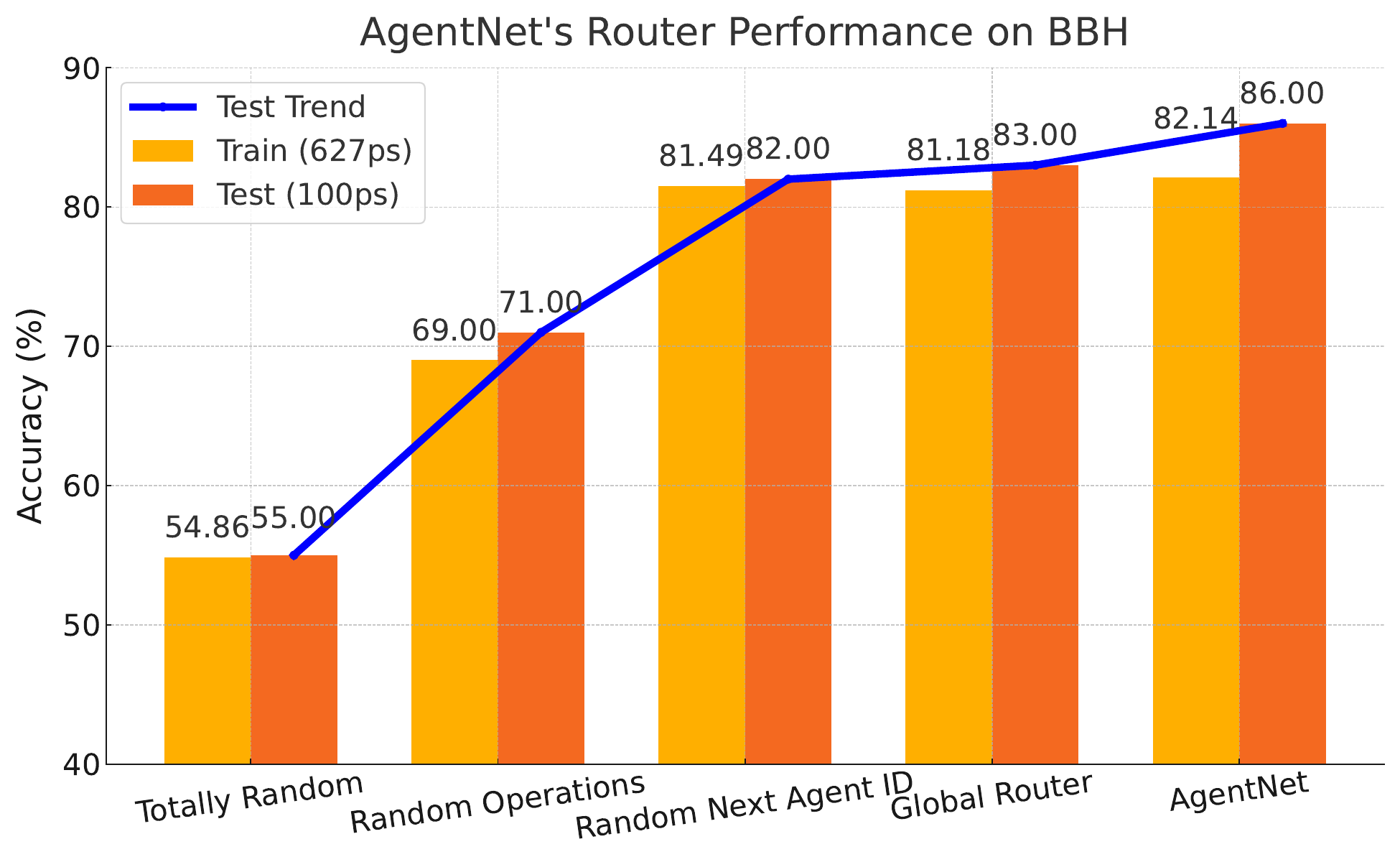}  % 替换成你的图像路径
\vspace{-0.3cm}
\caption{AgentNet's Router Performance on the BBH (Backbone: gpt-4o-mini)}
\label{fig:router}
\end{minipage}
\vspace{-0.2cm}
\end{figure}

\vspace{-0.5cm}

\begin{figure}[h!]
\centering
\begin{minipage}[t]{0.48\textwidth}
\vspace{+0.3cm}
\paragraph{Impact of Evolution Phase}
Results in Table~\ref{tab:warmup} clearly indicate that AgentNet significantly improves performance compared to the non-evolution baseline. On the MATH task, AgentNet achieves a score of 85.00 versus 77.86. For the function-calling task, performance improves notably from 23.00 to 32.00. On the BBH task, accuracy rises from 76\% to 86\%, demonstrating the impact of evolution phase for AgentNet.

\end{minipage}
\hfill
\begin{minipage}[t]{0.48\textwidth}
\begin{table}[H]
\centering
% \scriptsize
\vspace{-0.3cm}
\caption{Performance Comparison of AgentNet vs. Without evolution (Backbone: gpt-4o-mini)}
\vspace{+0.2cm}
\label{tab:warmup}
\renewcommand{\arraystretch}{1.1}
\rowcolors{2}{gray!10}{white}
\begin{tabular}{c|c|c|c}
\hline
\rowcolor{lightblue} \textbf{3 Agents} & \textbf{MATH} & \textbf{API-Bank} & \textbf{BBH} \\
\cmidrule(lr){2-4}
{} & \textbf{Acc(\%)} & \textbf{Acc(\%)} & \textbf{Acc(\%)} \\
\hline
w/o evolution & 77.86 & 23.00 & 76.00 \\
AgentNet & 85.00 & 32.00 & 86.00 \\
\hline
\end{tabular}
\end{table}
\end{minipage}
\vspace{-0.1cm}
\end{figure}

These results confirm that AgentNet’s adaptive learning during the evolution phase effectively enhances agent specialization and task performance, demonstrating its essential role in the system’s optimization and overall efficiency.

% \vspace{-0.3cm}

\subsection{Case Studies}
This case study is presented to illustrate the differences between the two methods, based on results obtained using GPT-4o-mini on the BBH dataset. The left image shows the trajectory produced by the ReAct method, while the right image illustrates the trajectory generated by AgentNet. In the case of ReAct, the lack of collective reasoning results in an incorrect response after a single-step inference, highlighting the limitations of the method in handling tasks that require more complex reasoning. In contrast, AgentNet uses a multi-step workflow where agents without the necessary expertise are bypassed, while those with the relevant skills divide the task into smaller steps, leading to a more accurate final solution.

\begin{figure}[h]
    \centering
    \small
    % Left side (ReAct Framework)
    \begin{minipage}{0.45\textwidth}
        \fbox{\begin{minipage}[t]{\textwidth}
        \raggedright
        \textbf{\textcolor{blue}{Question:}} \\
        \textcolor{black}{Which sentence has the correct adjective order?} \\[0.5ex]
        \textbf{\textcolor{blue}{Options:}} \\
        \textbf{A.} old-fashioned circular green cardboard exercise computer \\
        \textbf{B.} old-fashioned green exercise circular cardboard computer \\[1ex]
        \textbf{\textcolor{orange}{ReAct}} \\[0.5ex]
        \textbf{\textcolor{blue}{Reasoning:}} 
        (B) was selected because it follows a common pattern in English, where opinion adjectives (e.g., "old-fashioned") precede color adjectives (e.g., "green") and shape adjectives (e.g., "circular"). \\[1ex]
        \textbf{\textcolor{blue}{Action:}} \textcolor{red}{(B)}
        \textbf{\textcolor{blue}{\quad \quad \quad \quad \quad \quad \quad \quad \quad \quad Ground Truth:(A)}}
        % \textit{\textcolor{green}{This failure occurred because the reasoning process did not correctly apply the standard adjective order rules, leading to an incorrect selection of (B).}}
        \end{minipage}}
        \caption{ReAct Response with Reasoning}
    \end{minipage}
    \hspace{0.02\textwidth}
    % Right side (AgentNet)
    \begin{minipage}{0.5\textwidth}
        \fbox{\begin{minipage}[t]{\textwidth}
        \raggedright
        \textbf{\textcolor{orange}{AgentNet (5 Agents)}} \\[0.5ex]
        \textbf{Agent 0:} \textcolor{blue}{Decision:} Forward \textcolor{gray}{Next Agent: Agent 1} \\[0.5ex]

        \textbf{Agent 1:} \textcolor{blue}{Decision:} Forward \textcolor{gray}{Next Agent: Agent 2} \\[0.5ex]

        \textbf{Agent 2:} \textcolor{blue}{Decision:} Forward \textcolor{gray}{Next Agent: Agent 3} \\[0.5ex]

        \textbf{Agent 3:} \textcolor{blue}{Decision:} Split \\
        \textcolor{gray}{Executable: Adjective order analysis} \\
        \textcolor{gray}{Delegate: Final answer to another agent} \\
        \textcolor{gray}{Findings:}\\ 
        \quad \textcolor{gray}{Correct order: Quantity, opinion, size, age, shape, color, origin, material, purpose} 
        \quad \textcolor{gray}{Correct Option: (A)} \\[0.5ex]
        \textcolor{gray}{Next Agent: Agent 4}\\

        \textbf{Agent 4:} \textcolor{blue}{Decision:} Execute \textcolor{gray}{Action: Confirmed option (A)} \\[0.5ex]

        \textbf{\textcolor{blue}{Final Outcome:}} \textcolor{green}{(A)} \textbf{\textcolor{blue}{\quad \quad \quad \quad \quad \quad \quad \quad \quad Ground Truth:(A)}}
        \end{minipage}}
        \caption{AgentNet Task Breakdown}
    \end{minipage}
\end{figure}

\section{Analysis}
\subsection{Scalability and Robustness of the System}
\begin{figure*}[h]
    \centering
    \includegraphics[width=0.95\linewidth]{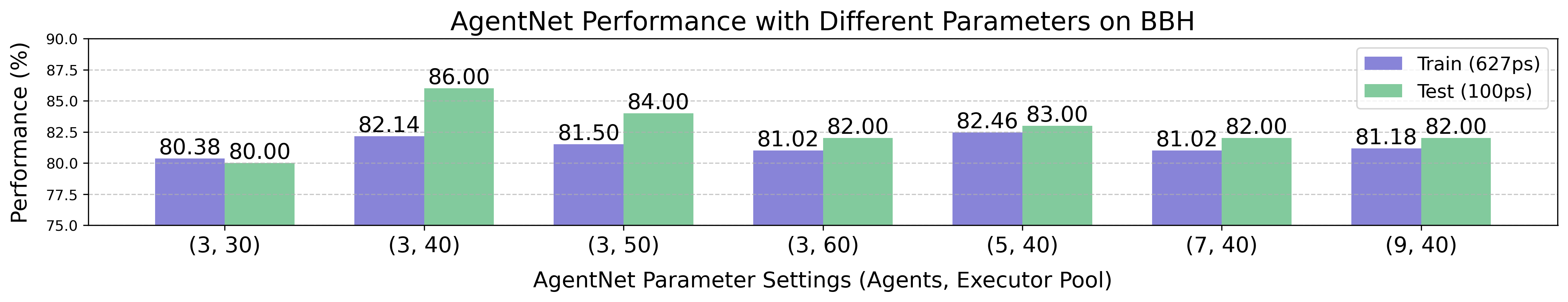}
    % \vspace{-0.2cm}
    \caption{AgentNet Performance with Different Net Parameters. Experiments were conducted with routers without pool limit, where (A, B) represents A as the number of agents and B as the upper limit of the executor pool, with performance evaluated on the BBH.}
    \label{fig:Parameters}
    % \vspace{-0.3cm}
\end{figure*}

Based on the experimental results as illustrated in Figure ~\ref{fig:Parameters}, we observed that both training and testing performance improve slightly as the number of agents and executor pool limit increase.   However, the improvements are incremental, with diminishing returns as the system scales up.   Specifically, performance in the training phase increased from 80.38 for 3 agents and 30 executors to 81.18 for 9 agents and 40 executors.   In the testing phase, performance fluctuated between 80 and 86, with the highest performance seen in configurations with 40 executors.   

These results suggest that AgentNet’s decentralized coordination model allows for gradual performance improvement as resources are added.   This indicates that AgentNet can scale effectively, enhancing performance without the dramatic bottlenecks commonly seen in centralized systems.   Additionally, while performance benefits from increased resources, the marginal gains suggest there may be an optimal point where resource allocation reaches its most efficient balance.  The experiment demonstrates that AgentNet’s design, which dynamically adjusts agent connections and executor pool sizes, effectively supports scalable and adaptable multi-agent systems with a high degree of fault tolerance.

\subsection{Autonomous Specialization of Agents}

\begin{figure*}[h]
    \centering
    \includegraphics[width=0.99\linewidth]{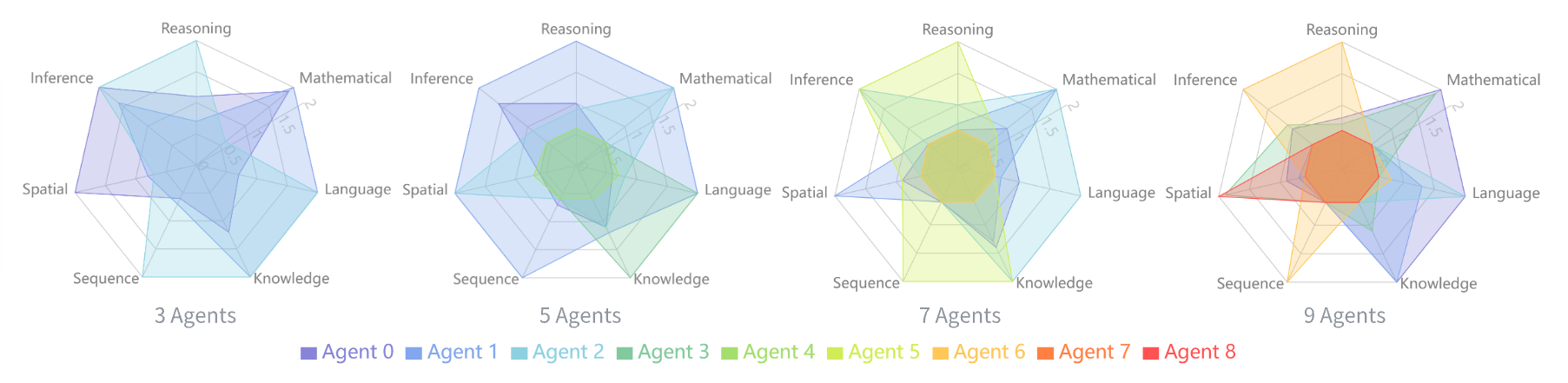}
    % \vspace{-0.2cm}
    \caption{Autonomous Specialization under Different Agent Sets.  The upper limit of the executor pool is fixed at 40, regardless of the number of agents.  The experiment was run on the BBH, and the images show the final ability scores of the agents after training.}
    \label{fig:all_agents}
    % \vspace{-0.3cm}
\end{figure*}

Based on the observed results in Figure ~\ref{fig:all_agents}, the experiments demonstrate that AgentNet's multi-agent system can naturally specialize agents in a decentralized environment. With varying numbers of agents and a fixed executor pool of 40 pieces, the ability scores across different tasks such as reasoning, language, knowledge, and sequence showed significant variation. As the number of agents increased, specialization became more evident, particularly in complex tasks, with certain agents excelling in specific areas while others focused on different abilities. This highlights AgentNet's capacity to dynamically refine agent expertise and optimize performance in a decentralized, task-driven system.

Notably, with 3 agents, the abilities were more evenly distributed across the agents, but as the number of agents increased, specialization became more evident, especially in tasks like knowledge and sequence, where specific agents showed notable proficiency.  In the (5, 40) configuration, the specialization became clearer, with some agents excelling in certain abilities, while others lagged behind in different tasks.  As we scaled up to 7 and 9 agents, the system displayed even greater specialization, especially in complex tasks, demonstrating AgentNet's ability to allow for dynamic expertise refinement.  This confirms that AgentNet supports the hypothesis of autonomous specialization within a decentralized, task-driven multi-agent system, where agents evolve to optimize task performance independently, without the need for a central controller.

\subsection{Evolution of Agents Networks}

\begin{figure*}[h]
    \centering
    \includegraphics[width=\linewidth]{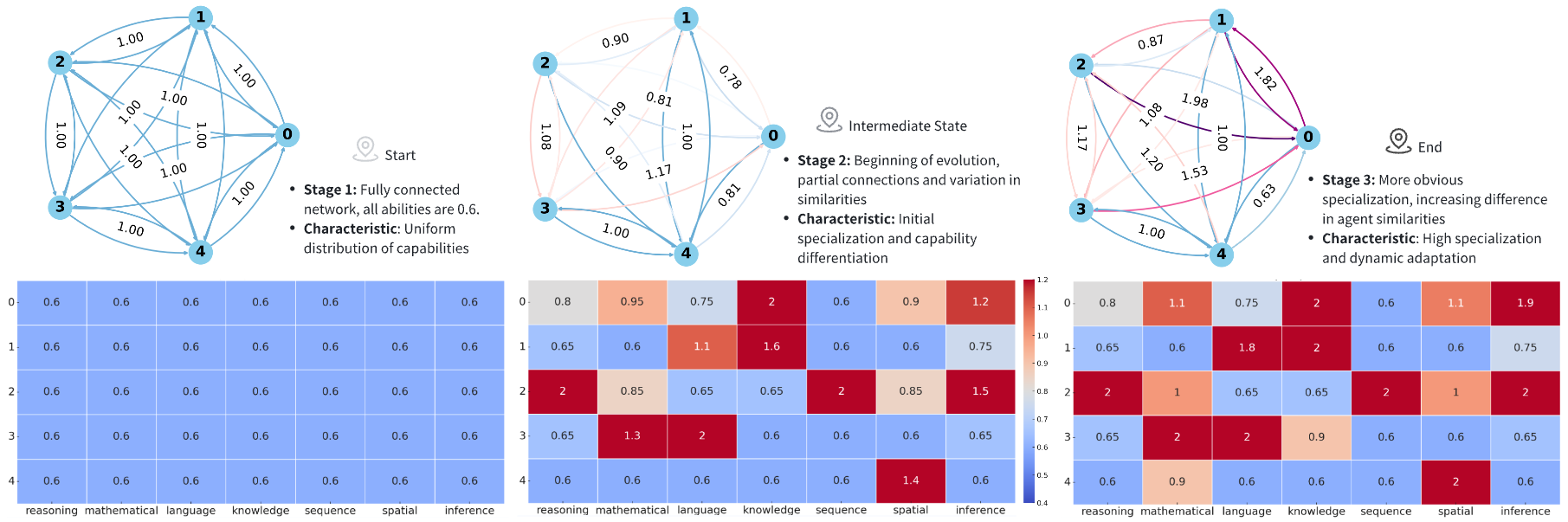}
    % \vspace{-0.2cm}
    \caption{Evolution Example of Agents Networks.}
    \label{fig:Evolution}
    % \vspace{-0.3cm}
\end{figure*}

The evolution of the agent network in our experiment is illustrated in Figure \ref{fig:Evolution}, which demonstrates the transition of a multi-agent system composed of 5 agents running on the BBH (627 pieces) benchmark. The figure captures the network at three key stages: the initial state, an intermediate state, and the final evolved state.

In the initial state, the network is fully connected with uniform connection values of 1.00, indicating equal capabilities among all agents. At this stage, there is no specialization, and all agents are equally equipped to handle tasks.

As the network evolves, agents begin to specialize, and connection values vary, reflecting the strength of collaboration. Stronger connections indicate tighter cooperation, while weaker ones suggest less interaction. This evolution shows how agents naturally adapt and form more efficient collaboration patterns.

By the final stage, the network exhibits clear specialization, with agents taking on distinct roles. The connection values further emphasize the growing cooperation between specialized agents, improving task performance. This progression demonstrates the effectiveness of decentralized coordination, where evolving collaboration enhances task allocation, scalability, and fault tolerance.

\section{Limitations and Future Work}
Despite AgentNet implementing a fully distributed, adaptive learning multi-agent system (MAS) with dynamic task allocation, several important limitations remain that require further exploration in future work.   

One key challenge is how to improve task performance in heterogeneous agent environments.   In real-world applications, agents often vary significantly in terms of model capabilities, workflow structures, tools, and available data.   The impact of such heterogeneity on AgentNet’s performance, especially in terms of task coordination and resource allocation, remains an open question.   Understanding how to adapt the system to handle such variations efficiently will be crucial for its scalability and effectiveness in complex environments.

Secondly, the decision-making process of the router within each agent, particularly in relation to exploration and discovery, requires more in-depth study.   Currently, the router selects agents from a relatively small pool of predefined candidates.   However, in larger-scale systems involving hundreds or potentially thousands of agents, the challenge of accurately identifying the most suitable agent for task delegation becomes significantly more complex.   This problem is further compounded in heterogeneous settings, where different agents may possess distinct strengths and weaknesses.   To address this, future research could focus on developing more sophisticated routing mechanisms that can autonomously identify and delegate tasks to the most appropriate agents, even in large and diverse agent pools.   

Additionally, a promising direction for future work involves designing incentives that encourage the router to explore agents beyond the predefined candidate set.   By enabling AgentNet to dynamically discover new agents or specialized capabilities, such an approach would enhance its adaptability and scalability, ultimately improving the system’s overall performance and autonomy.

\section{Conclusion}
% \vspace{-0.2cm}
In conclusion, AgentNet provides an effective approach to addressing the limitations of traditional centralized multi-agent systems. With its decentralized architecture, dynamic task allocation, and adaptive learning mechanisms, AgentNet improves scalability, fault tolerance, and task efficiency in collaborative environments. Its privacy-preserving features further ensure secure cooperation across organizations. Our experimental results highlight the advantages of this approach, demonstrating improvements in task efficiency, adaptability, and specialization. AgentNet offers a practical framework for developing more flexible and secure multi-agent systems in dynamic, real-world settings.

\section*{Acknowledgement}
The authors gratefully acknowledge the generous support provided by Bytedance, which has been instrumental in enabling the successful execution of this research. Their sponsorship significantly contributed to the advancement and completion of this work.

\section*{Acknowledgement}
This research was supported by Bytedance through a sponsored project that facilitated the execution and completion of this work.

\section*{Ethics Statement}
This study did not involve human participants, animal subjects, or the use of personal data. All datasets and benchmarks employed are publicly available and used in accordance with their respective licenses. Therefore, no ethics approval was required.

\bibliographystyle{unsrtnat}
\bibliography{references}  %%% Uncomment this line and comment out the ``thebibliography'' section below to use the external .bib file (using bibtex) .

\newpage
% 添加大标题 APPENDIX
% \bigappendix
\appendix





% \end{thebibliography}

\end{document}